\documentclass[reprint,aps,prb,citeautoscript,showpacs]{revtex4-1}
\usepackage[utf8]{inputenc}
\usepackage{graphicx}
\usepackage{amsmath}
\usepackage{amssymb}
\usepackage{verbatim}
\usepackage[]{hyperref}
\usepackage{siunitx}
\usepackage{booktabs}

\newcommand{\bra}[1]{\langle #1 |}
\newcommand{\ket}[1]{| #1 \rangle}

\begin{document}

\pacs{71.15.Dx, 71.15.Qe, 73.22.-f}

\begin{abstract}
Calculating the quasiparticle (QP) band structure of two-dimensional (2D) materials within the GW self-energy approximation has proven to be a rather demanding computational task. The main reason is the strong $\mathbf{q}$-dependence of the 2D dielectric function around $\mathbf{q} = \mathbf{0}$ that calls for a much denser sampling of the Brillouin zone than is necessary for similar 3D solids. Here we use an analytical expression for the small $\mathbf{q}$-limit of the 2D response function to perform the BZ integral over the critical region around $\mathbf{q} = \mathbf{0}$. This drastically reduces the requirements on the $\mathbf{q}$-point mesh and implies a significant computational speed-up. For example, in the case of monolayer MoS$_2$, convergence of the $G_0W_0$ band gap to within $\sim\SI{0.1}{eV}$ is achieved with $12\times 12$ $\mathbf{q}$-points rather than the $36\times 36$ mesh required with discrete BZ sampling techniques. We perform a critical assessment of the band gap of the three prototypical 2D semiconductors MoS$_2$, hBN, and phosphorene including the effect of self-consistency at the GW$_0$ and GW level. The method is implemented in the open source GPAW code. 
\end{abstract}

\title{Efficient many-body calculations of 2D materials using exact limits for the screened potential: Band gaps of MoS$_2$, hBN, and phosphorene}
\author{Filip A. Rasmussen}
\author{Per S. Schmidt}
\author{Kirsten T. Winther}
\author{Kristian S. Thygesen}
\affiliation{Center for Atomic-scale Materials Design (CAMD), Department of Physics and Center for Nanostructured Graphene (CNG), Technical University of Denmark, DK-2800 Kongens Lyngby, Denmark}
\date{\today}

\maketitle

\section{Introduction}
The past few years have witnessed an explosion in research on atomically thin two-dimensional (2D) materials. Of particular interest are the 2D semiconductors including the family of transition metal dichalcogenides, which have been found to exhibit a number of unique opto-electronic properties\cite{mak_atomically_2010, splendiani_emerging_2010, zeng_optical_2013, zhang_direct_2014,wang_electronics_2012,britnell_strong_2013, bernardi_extraordinary_2013}. For understanding and predicting these properties the electronic band structure of the material is of fundamental importance. The GW method\cite{aulbur_quasiparticle_1999, aryasetiawan_gw_1998}, introduced by Hedin\cite{hedin_new_1965} in 1965 and first applied to real solids in an ab-initio framework by Hybertsen and Louie\cite{hybertsen_first-principles_1985} and Godby, Sham, and Schlüter\cite{godby_accurate_1986}, has become the ``gold standard'' for calculating quasi-particle (QP) band structures. Over the years its performance has been thoroughly established for bulk materials\cite{shishkin_implementation_2006, kotani_all-electron_2002, marini_yambo:_2009} and more recently also for molecules\cite{rostgaard_fully_2010, caruso_self-consistent_2013, bruneval_$gw$_2009, blase_first-principles_2011}. In comparison, critical assessments of the accuracy and numerical convergence of GW calculations for 2D materials are rather scarce\cite{ismail-beigi_truncation_2006, huser_how_2013, qiu_erratum:_2015,rasmussen_computational_2015}. Nevertheless these studies have shown that (i) it is essential to use a truncated Coulomb interaction to avoid long range screening between periodically repeated layers which reduces the QP band gap, and (ii) when a truncated Coulomb interaction is used, the convergence of the GW calculation with respect to the number of $\mathbf{k}$-points becomes much slower than is the case for similar bulk systems.   

The slow $\mathbf{k}$-point convergence of the GW band structure is directly related to the nature of electronic screening in 2D which is qualitatively different from the well known 3D case.\cite{keldysh_coulomb_1979, cudazzo_dielectric_2011} Specifically, while the dielectric function, $\varepsilon(\mathbf{q})$, of bulk semiconductors is approximately constant for small wave vectors, the dielectric function of a 2D semiconductor varies sharply as $\mathbf{q}\to\mathbf{0}$.\cite{ismail-beigi_truncation_2006, huser_how_2013} As a consequence, the number of $\mathbf{q}$-points required to obtain a proper sampling of the screened interaction $W(\mathbf{q})$ over the Brillouin zone (BZ) is much higher for the 2D material than what would be anticipated from the 3D case. For example, the band gap of bulk MoS$_2$ is converged to within $\sim\SI{0.1}{eV}$ with an in-plane $\mathbf{k}$-point grid of $12\times 12$ while the same accuracy for monolayer MoS$_2$ requires a grid of $36 \times 36$ when standard BZ sampling schemes are applied.

Importantly, supercell calculations only describe the characteristic features of screening in 2D materials correctly when the unphysical screening between the periodically repeated layers is removed, e.g. using a truncated Coulomb interaction. Without this, the dielectric function does not approach unity for $\mathbf{q}\to 0$ and the band gap can be significantly underestimated (by around \SI{0.5}{eV} for MoS$_2$ with \SI{10}{Å} vacuum\cite{huser_how_2013}) as a result of over screening. Since in this case, the screening is really intermediate between 3D and 2D, the GW gap shows faster convergence with $k$-points than is observed using a truncated Coulomb interaction. This is presumably the reason why most earlier GW calculations for 2D semi-conductors have been performed with $k$-point grids ranging from $6\times 6$ to $15\times 15$ which are much too coarse for describing the truly isolated 2D material. 

Here we show that the slow $\mathbf{k}$-point convergence of the GW self-energy in 2D materials can be alleviated by performing the BZ integral of $W(\mathbf{q})$ analytically in the critical region around $\mathbf{q} = \mathbf{0}$ where $\varepsilon(\mathbf{q})$ varies most strongly. The analytical expression for $W(\mathbf{q})$ is obtained from a lowest order expansion in $\mathbf{q}$ of the head, $\chi^0_\mathbf{00}(\mathbf{q})$, and wings, $\chi^0_\mathbf{0G}(\mathbf{q})$, of the non-interacting density response function. This simple scheme reduces the required number of $\mathbf{q}$-points by an order of magnitude without loss of accuracy. 

\section{The GW self-energy}
We split the GW self-energy into the exchange and correlation part, respectively. The former does not present particular problems in 2D materials and is calculated using a Wigner-Seitz truncated Coulomb interaction as described elsewhere\cite{sundararaman_regularization_2013}. In a plane wave expansion the correlation part of the self-energy, evaluated for an electronic state $\ket{n\mathbf{k}}$, takes the general form\cite{hybertsen_electron_1986}
\begin{equation}
\begin{split}
\bra{n\mathbf{k}} \Sigma^c(\omega) \ket{n\mathbf{k}} = \frac{1}{(2\pi)^3}\int_\text{BZ} d\mathbf{q} \sum_{\mathbf{GG}'} \frac{i}{2\pi} \int_{-\infty}^\infty d\omega' \overline{W}_{\mathbf{GG}'}(\mathbf{q}, \omega') \\
\times \sum_m \frac{[\rho_{n,\mathbf{k}}^{m,\mathbf{k} + \mathbf{q}}(\mathbf{G})] [\rho_{n\mathbf{k}}^{m,\mathbf{k}+\mathbf{q}}(\mathbf{G}')]^*}{\omega + \omega' - \epsilon_{m\mathbf{k} + \mathbf{q}} - i\eta\,\text{sgn}(\epsilon_{m\mathbf{k} + \mathbf{q}}-\mu)}, \label{eq:self-energy}
\end{split}
\end{equation}
where $m$ runs over all electronic bands, $\epsilon_{m\mathbf{k}+\mathbf{q}}$ are single particle energies, $\mu$ is the chemical potential and $\eta$ is a broadening parameter. The pair densities are defined as $\rho_{n\mathbf{k}}^{m\mathbf{k} + \mathbf{q}}(\mathbf{G}) = \bra{n\mathbf{k}} e^{i(\mathbf{G} + \mathbf{q}) \cdot \mathbf{r}} \ket{m\mathbf{k} + \mathbf{q}}$, and $\overline{W}_{\mathbf{GG}'}(\mathbf{q}, \omega)$ is the correlation part of the dynamical screened potential given by
\begin{equation}
\overline{W}_{\mathbf{GG}'}(\mathbf{q}, \omega) = v_\mathbf{G}(\mathbf{q}) \left[\varepsilon^{-1}_{\mathbf{GG}'}(\mathbf{q}, \omega) - \delta_{\mathbf{GG}'}\right], \label{eq:screened-potential}
\end{equation}
where $v_\mathbf{G}(\mathbf{q})=4\pi/|\mathbf{G}+\mathbf{q}|^2$ is the Coulomb interaction. In most implementations the BZ integral is evaluated numerically with a standard quadrature method using a regular $\mathbf{q}$-point grid matching the $\mathbf{k}$-point grid of the ground state DFT calculation. Since the screened potential, Eq.~(\ref{eq:screened-potential}), diverges for $\mathbf{G} = 0,\;\mathbf{q} = \mathbf{0}$ (for bulk materials) this point is generally handled separately, so the integral may be written
\begin{equation}
 \int_\text{BZ} d\mathbf{q} \mathcal{S}(\mathbf{q}, \omega) \to \frac{\Omega}{N_\mathbf{q}}\sum_{\mathbf{q}_n \neq \mathbf{0}}\mathcal{S}(\mathbf{q}_n, \omega) + \int_{\Omega_0} d\mathbf{q} \mathcal{S}(\mathbf{q}, \omega), \label{eq:sigma_num_int}
\end{equation}
where $\mathcal{S}(\mathbf{q}, \omega)$ denotes the entire integrand, $\Omega$ is the volume of the BZ, $N_\mathbf{q}$ is the total number of $\mathbf{q}$-points in the grid and $\Omega_0$ denotes a small region around $\mathbf{q} = \mathbf{0}$. For bulk systems $\Omega_0$ is normally defined as a sphere centered at $\mathbf{q} = \mathbf{0}$. For a 2D material, the BZ integral in Eq.~(\ref{eq:self-energy}) reduces to a 2D integral with in-plane sampling of $\mathbf{q}$, and $\Omega_0$ represents an area. We now focus on how to calculate the contribution to the integral around the special point $\mathbf{q} = \mathbf{0}$ in the 3D versus the 2D case.  

%






Within the random phase approximation (RPA) the dielectric matrix is given by 
\begin{equation}\label{eq:epsilon}
  \varepsilon_{\mathbf{GG}'}(\mathbf{q}, \omega) = \delta_{\mathbf{GG}'} - v_\mathbf{G}(\mathbf{q})\chi^0_{\mathbf{GG}'}(\mathbf{q}, \omega).
\end{equation}
For a solid with a finite band gap it can be shown that the head of the non-interacting response function $\chi^0_\mathbf{00}(\mathbf{q},\omega) \propto q^2$ for small $q$\cite{pick_microscopic_1970}. Since $v_\mathbf{0}(\mathbf{q}) = 4\pi/q^2$ it follows that in 3D the head of the dielectric function $\varepsilon_\mathbf{00}(\mathbf{q}, \omega)$ converges to a finite value $>1$ when $q\to 0$. Moreover, this value is typically a reasonable approximation to $\varepsilon_\mathbf{00}(\mathbf{q}, \omega)$ in a relatively large region around $\mathbf{q} = \mathbf{0}$. This means that in the BZ integration in Eq.~(\ref{eq:self-energy}) around the singular point $\mathbf{G}=\mathbf{G}’=\mathbf{q}=\mathbf{0}$ all factors, except $4\pi/q^2$, can be assumed to be constant and the integral can be performed analytically over a sphere centred at $\mathbf{q} = \mathbf{0}$\cite{hybertsen_first-principles_1985} or numerically on a fine real space grid\cite{deslippe_berkeleygw:_2012}. 

For GW calculations of 2D materials performed with periodic boundary conditions in the out-of-plane direction, the direct use of Eq.~(\ref{eq:self-energy}) leads to significant overscreening due to the interaction between the repeated images\cite{huser_how_2013}. One way of dealing with this is to subtract the artificial interlayer contribution calculated from a classical electrostatic model\cite{freysoldt_screening_2008}. A more direct way of avoiding the spurious interactions is to truncate the Coulomb potential in the direction perpendicular to the layers. Thus in Eqs. (\ref{eq:epsilon}) and (\ref{eq:screened-potential}), $v_\mathbf{G}(\mathbf{q})$ should be replaced by\cite{rozzi_exact_2006, ismail-beigi_truncation_2006} 
\begin{equation}\label{eq:2Dcoulomb}
  v^\text{2D}_\mathbf{G}(\mathbf{q}_\parallel) = \frac{4\pi}{|\mathbf{q}_\parallel + \mathbf{G}|^2} \left[ 1 - e^{-|\mathbf{q}_\parallel + \mathbf{G}_\parallel| L / 2} \cos(|G_z| L / 2) \right],
\end{equation}
where only in-plane $\mathbf{q}$ are considered. $L$ is the length of the unit cell in the non-periodic direction and the truncation distance is set to $L/2$, which simplifies the expression. In the long wavelength limit, $\mathbf{G} = \mathbf{0}$, $\mathbf{q}_\parallel \to \mathbf{0}$, the truncated interaction becomes $v^\text{2D}_\mathbf{0}(q_\parallel)\approx \frac{2\pi L}{q_\parallel}$. We see that the $\mathbf{q}=\mathbf{0}$ divergence in the truncated Coulomb potential is reduced by a power of $\mathbf{q}$ compared to that of the full Coulomb interaction. As will be shown in the following this has important consequences for the form of the screened interaction. However, before presenting the form of the screened interaction of a 2D semiconductor evaluated using the full expression for the response function and truncated Coulomb interaction, it is instructive to consider a simplified description of the 2D material. 

Let us consider a strict 2D model of a homogeneous and isotropic semiconductor. In the small $\mathbf{q}$ limit, the density response function takes the form $\chi_0(\mathbf{q})=\alpha_\text{2D} q^2$ where $\alpha_\text{2D}$ is the 2D polarizability\cite{cudazzo_dielectric_2011}. Using that the 2D Fourier transform of $1/r$ equals $2\pi/q$, the leading order of the dielectric function becomes 
\begin{equation}\label{eq:2Deps}
\varepsilon^{2D}(\mathbf{q}) \approx 1 + 2\pi \alpha_\text{2D} q.
\end{equation}
\begin{figure}
  \includegraphics[width=\columnwidth]{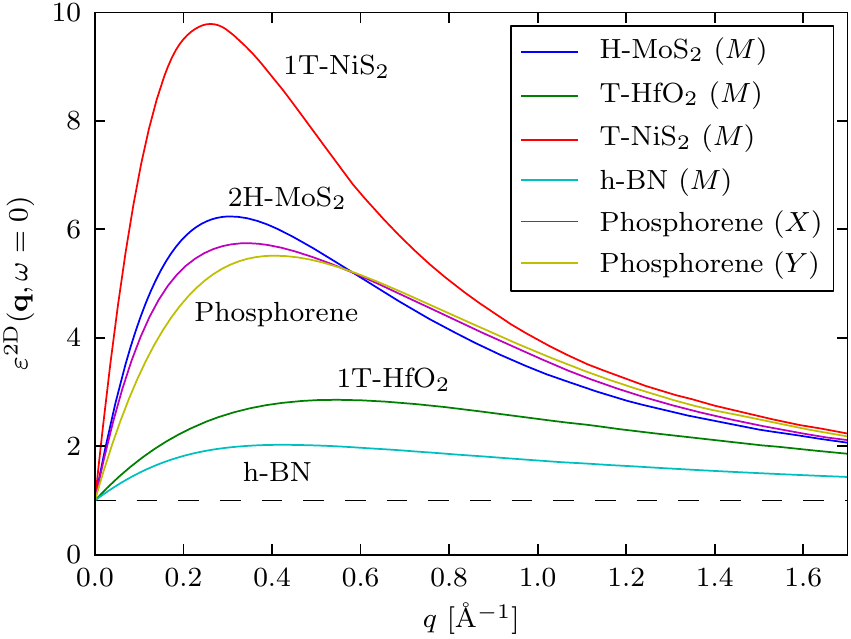}
  \caption{(Color online) Static macroscopic dielectric functions of a representative set of 2D semiconductors as a function of $\mathbf{q}$ along the $\Gamma \to M$ direction for the hexagonal structures and along the path from $\Gamma$ to $X$ or $Y$ in the case of phosphorene.}
  \label{fig:2d_eps}
\end{figure}
Some examples of \emph{macroscopic} dielectric functions for a representative set of 2D semiconductors are shown in Figure~\ref{fig:2d_eps} (see Ref.~\onlinecite{huser_how_2013} for a precise definition of this quantity). The linear form (\ref{eq:2Deps}) is clearly observed in the small $\mathbf{q}$ regime. Importantly, if we use the same strategy for evaluating the BZ integral in Eq.~(\ref{eq:self-energy}) as in 3D, i.e. assuming $\epsilon^{-1}(\mathbf{q})$ to be a slowly varying function and evaluating it on the discrete $\mathbf{q}$-point grid, we obtain zero contribution for the $\mathbf{q} = 0$ term because $1/\epsilon^{2D}-1=0$ for $\mathbf{q}=0$, see Eq. \eqref{eq:screened-potential}. On the other hand, it is clear that the screened interaction takes the form $\overline{W}^{2D}(\mathbf{q}) = -4\pi^2 \alpha_\text{2D}/(1 + 2\pi \alpha_\text{2D} q)$ for small $\mathbf{q}$. In particular, $\overline{W}^{2D}(\mathbf{q})$ takes a finite value for $\mathbf{q}=0$ which is qualitatively different from the 3D case where $\overline{W}(\mathbf{q})$  diverges for $\mathbf{q} \to \mathbf{0}$.

In Appendix~\ref{sec:appendix} we show, following an analysis similar to that of Ref.~\onlinecite{freysoldt_dielectric_2007} adapted to the case of  a truncated Coulomb interaction, that for a general non-isotropic 2D material, the small $\mathbf{q}_\parallel$ limit of the head of the screened potential takes the form
\begin{equation}
\begin{split}
  \overline{W}_\mathbf{00}(\mathbf{q}_\parallel) ={}& - \left(\frac{4\pi (1 - e^{-|\mathbf{q}_\parallel| L/2})}{|\mathbf{q}_\parallel|} \right)^2 \\
 & \times \frac{\hat{\mathbf{q}_\parallel} \cdot \mathsf{A}\hat{\mathbf{q}_\parallel}}{1 + 4\pi (1 - e^{-|\mathbf{q}_\parallel| L/2}) \hat{\mathbf{q}_\parallel} \cdot \mathsf{A} \hat{\mathbf{q}_\parallel}},\label{eq:exact}
\end{split}
\end{equation}
where $\hat{\mathbf{q}_\parallel} = \mathbf{q}_\parallel / |\mathbf{q}_\parallel|$ and $\mathsf{A}$ is a second rank tensor which also depends on the frequency. We see that we have $\overline{W}_\mathbf{00}(\mathbf{q}_\parallel = \mathbf{0}) = -(2\pi L)^2 \hat{\mathbf{q}_\parallel} \cdot \mathsf{A} \hat{\mathbf{q}_\parallel}$. The expression $\hat{\mathbf{q}_\parallel} \cdot \mathsf{A}\hat{\mathbf{q}_\parallel}$ is closely related to the slope of the dielectric function and the 2D polarizability but includes local field effects. In addition to Eq.~(\ref{eq:exact}) there are similar expressions for the wings and body of the screened interaction, see Eq.\eqref{eq:Wheadfull} to Eq.\eqref{eq:Wbodyfull}. These expressions must be integrated over the $\Omega_0$-region, that we now define as the primitive cell in the 2D BZ that surrounds the $\mathbf{q}_\parallel =\mathbf{0}$ point. The expression is simplified to one that can be integrated analytically as shown in Appendix \ref{sec:appendix}. 
 

\begin{figure}[h!t]
  \includegraphics[width=\columnwidth]{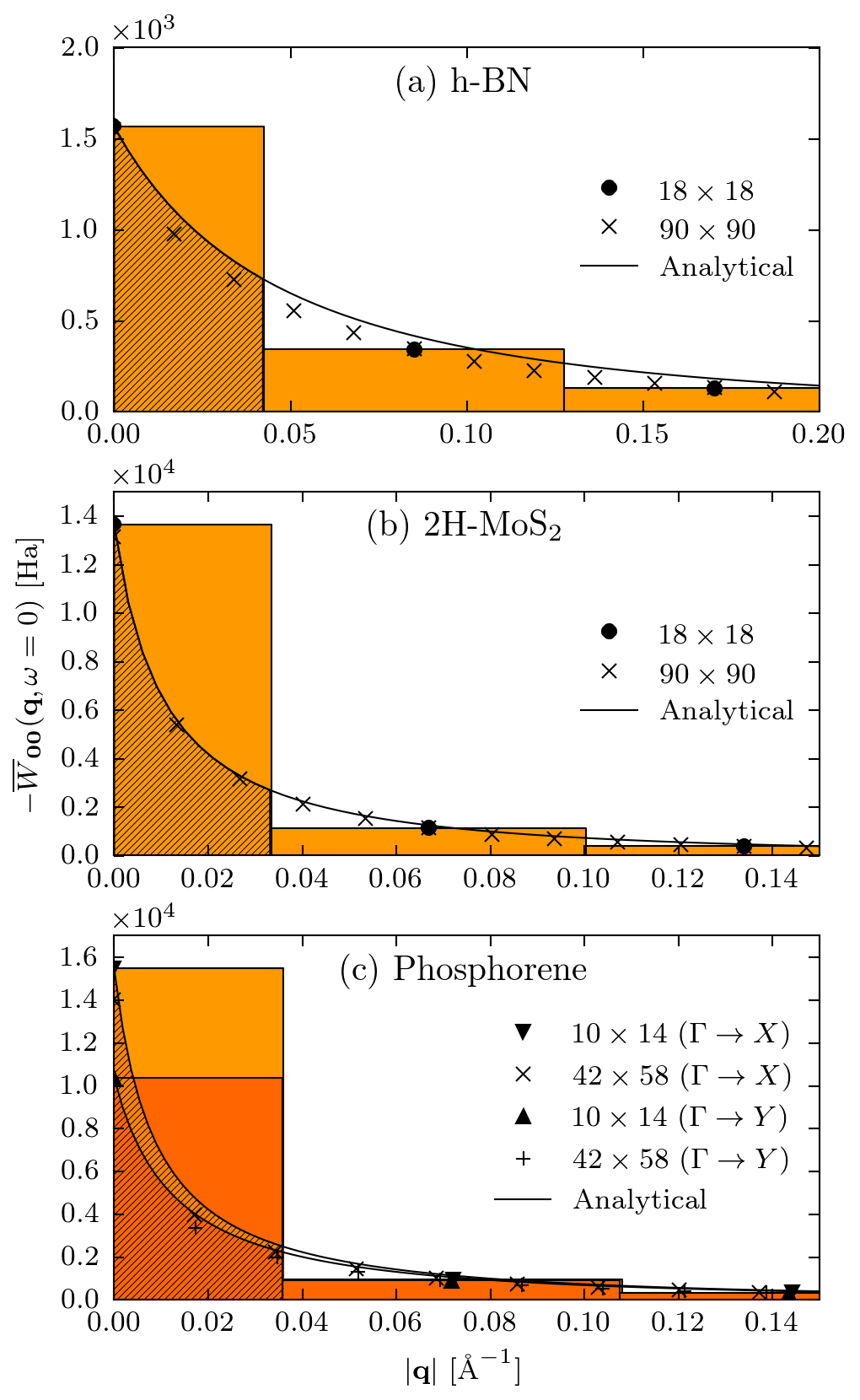}
  \caption{(Color online) The head of the static component of the screened potential (subtracted the bare interaction) of monolayer a) h-BN b) MoS$_2$ and c) phosphorene as a function of $\mathbf{q}$ along the $\Gamma\to M$ direction or $\Gamma\to X$ and $\Gamma\to Y$ in the case of phosphorene. The crosses are the numerical values obtained on a fine $\mathbf{q}$-point grid while the circles or triangles represent the values obtained on a coarse $\mathbf{q}$-point grid. The bars represent a simple numerical approximation to the BZ integral of $\overline{W}_{00}(\mathbf{q})$ performed on the coarse $\mathbf{q}$-point grid. The value of the screened potential for $\mathbf{q} = \mathbf{0}$ is set to the analytical result Eq.~(\ref{eq:exact}). The full curve represents the analytical small $\mathbf{q}$ approximation, Eq.~(\ref{eq:exact}), and the hatched area shows its contribution to the integral.}
  \label{fig:W_qint}
\end{figure}

The full expression for $\overline{W}$ in Eq.~(\ref{eq:exact}) is therefore evaluated numerically on a discrete sub-grid, constructed as a Monkhorst-Pack grid within $\Omega_0$, and the simplified expression in Eq.~(\ref{eq:w_taylor}) is only used for $\mathbf{q}_\parallel = \mathbf{0}$ on the sub-grid. The limit of the integral is now given by the radius $r_{\Omega_0}$, defined as $\pi r^2_{\Omega_0} = \Omega_0 / N_{q_0}$, where $N_{q_0}$ is the number of grid points in the sub-grid. This approach ensures a smooth evaluation of $\overline{W}$, that converges fast with $N_{q_0}$. It is found to be necessary to have $N_{q_0}\approx 100$ when $\mathbf{q}_\parallel = \mathbf{0}$ is evaluated using Eq. ~(\ref{eq:w_taylor}) for both iso- and anisotropic materials where as $N_{\mathbf{q_0}}\approx 10^5$ is needed if the analytical correction at $\mathbf{q}_\parallel = \mathbf{0}$ is omitted. 

\subsection{Results}
To investigate how this method performs we have carried out test calculations for the three monolayers h-BN, MoS$_2$ and phosphorene, which have quite different dielectric functions as seen on Figure~\ref{fig:2d_eps}. h-BN is a large gap dielectric with low screening ability leading to a small slope of the dielectric function at $\mathbf{q} = \mathbf{0}$, while MoS$_2$ has a larger dielectric function and quite steep slope at $\mathbf{q} = \mathbf{0}$. Phosphorene has a dielectric function similar to MoS$_2$ in size and steepness but is anisotropic with slopes varying by $\sim 40\%$ between the two high symmetry directions, $\Gamma\to X$ and $\Gamma\to Y$. 

All the calculations were performed using the GPAW electronic structure code\cite{GPAW, GPAW_review, huser_quasiparticle_2013}. The structures used in the present calculations are relaxed with DFT using the PBE exchange-correlation (xc) functional\cite{PBE}. The resulting lattice constant for h-BN is \SI{2.504}{Å}, the in-plane lattice constant for MoS$_2$ is \SI{3.184}{Å} with a S-S distance of \SI{3.127}{Å}. For phosphorene the in-plane unit cell is \SI{4.630}{Å} by \SI{3.306}{Å}, the in-plane P-P-P angle is \SI{95.8}{\degree} and the layer thickness is \SI{2.110}{Å}. A convergence test with respect to the amount of vacuum between repeating periodic images was carried out and \SI{15}{Å} of vacuum was necessary for h-BN and Phosphorene where as only \SI{10}{Å} was needed for MoS$_2$. The PBE eigenvalues and wavefunctions were calculated with a plane wave basis cut-off energy of \SI{600}{eV} and used as input in the GW calculations. For the initial investigation of the $\mathbf{q}$-point convergence, the dielectric function and the correlation self-energy were calculated using a cutoff of \SI{50}{eV}. This cutoff is insufficient to ensure properly converged quasi-particle energies, but it is adequate to describe the trends related to the improved $\mathbf{q}$-point sampling relevant for this study. The following fully converged calculations were carried out using a $1/N_\text{pw}$ extrapolation to the complete basis set limit using cutoff energies of up to \SI{200}{eV}\cite{tiago_effect_2004, shih_quasiparticle_2010, klimes_predictive_2014}.

In Figure~\ref{fig:W_qint} we compare the analytical small $\mathbf{q}$ expression, Eq.~(\ref{eq:exact}), for the head of the screened potential $\overline{W}_\mathbf{00}(\mathbf{q})$ with the numerical values obtained using a fine and coarse $\mathbf{q}$-point sampling. In all the cases the $\mathbf{q} = \mathbf{0}$ value has been set to the analytical value. It is evident that the screened potential falls off quickly and thus for a coarse $\mathbf{q}$-point sampling the $\mathbf{q} = \mathbf{0}$ contribution to the integral is by far the largest and should therefore not be neglected. Similarly, using only the exact value in $\mathbf{q}=\mathbf{0}$ could also pose a problem as the contribution will be grossly overestimated due to the convex nature of potential. We note that the analytical expression follows the numerical results quite closely and is even accurate far away from the $\Gamma$-point -- for MoS$_2$ we have an almost perfect agreement for the points shown. Thus using the analytical limit within the region around $\mathbf{q}=\mathbf{0}$ is reasonable. We notice that the anisotropy of phosphorene makes $\overline{W}_\mathbf{00}(\mathbf{q})$ ill-defined at $\mathbf{q} = \mathbf{0}$ (different limit values depending on the direction of $\mathbf{q}$). For larger $\mathbf{q}$ the dielectric anisotropy becomes negligible. However, because of the relatively large weight of the $\mathbf{q} = \mathbf{0}$ contribution to the BZ integral, the anisotropy should be taken into account for accurate GW calculations.

We note that a similar approach to the treatment of the $\mathbf{q} = \mathbf{0}$ term of the screened potential was suggested in Ref.~\onlinecite{ismail-beigi_truncation_2006}. That particular method was based on fitting to an empirical expression for $\varepsilon(q)$ calculated from the value at a small but finite $\mathbf{q}$. The method outlined here is different in that the analytical expression for $\overline{W}(\mathbf{q})$ is obtained from a lowest order expansion of the head, $\chi^0_\mathbf{00}(q)$, and wings, $\chi^0_\mathbf{0G}(q)$, of the non-interacting density response function\cite{yan_linear_2011} and thus can be obtained without fitting or using empirical parameters. This also ensures that the effect of in-plane dielectric anisotropy is explicitly included.

\begin{figure}[h!t]
  \includegraphics[width=\columnwidth]{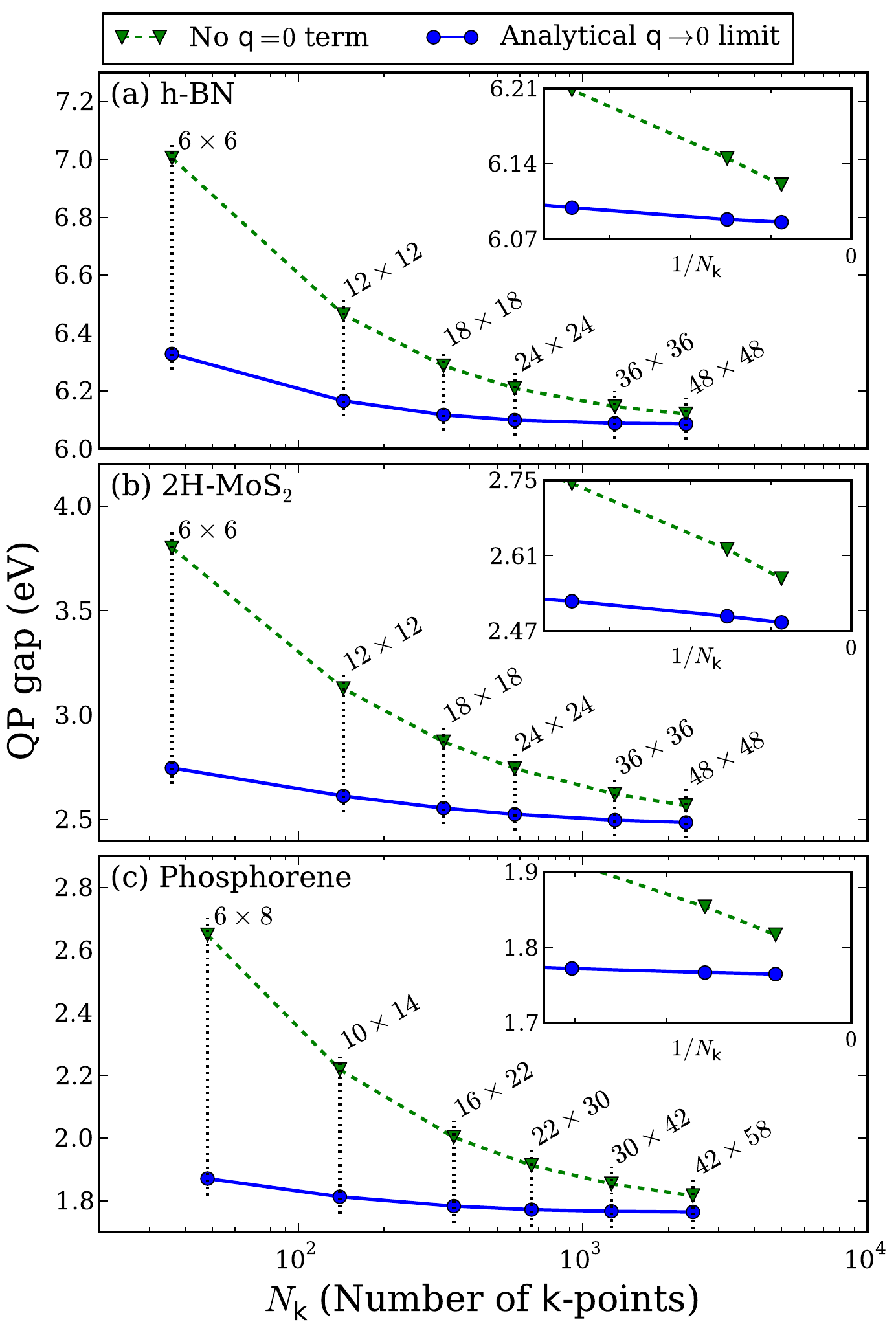}
  \caption{(Color online) The G$_0$W$_0$ quasi-particle band gap of monolayer (a) 2H-MoS$_2$ (b) h-BN and (c) phosphorene, calculated using two different treatments of the $\mathbf{q} = \mathbf{0}$ term in Eq.~(\ref{eq:self-energy}). The dashed (green) line shows the contribution obtained when the head and wing elements of the $\mathbf{q} = \mathbf{0}$ term are neglected corresponding to the standard treatment used for 3D systems. The solid (blue) line shows the contribution obtained when using the analytical results, Eq.~(\ref{eq:exact}), to perform the integral over the $\mathbf{q} = \mathbf{0}$ element. The insets shows the results for the largest $\mathbf{k}$-point grids on a reversed linear scale in $1/N_\mathbf{k}$. Notice the zero point is at the right side of the $x$-axis.}
  \label{fig:kpt-convergence}
\end{figure}

In Fig.~\ref{fig:kpt-convergence} we show the minimum QP band gap of monolayer h-BN, MoS$_2$ and phosphorene as a function of $1/N_\mathbf{k}$ where $N_\mathbf{k}$ is the total number of $\mathbf{k}$-points in the BZ sampling (the $\mathbf{q}$ point grid for the GW integration is the same as the $\mathbf{k}$-point grid used in DFT). We compare the results obtained using two methods: i) neglecting the $\mathbf{q} = \mathbf{0}$ contribution to head and wings of the screened potential and ii) evaluating Eq.~(\ref{eq:exact}) as described. It is clear that method i) in all cases underestimates the correlation self-energy due to the underestimation of the screening; In order to get the band gap converged to within $\sim\SI{0.1}{eV}$ one would have to use a $\mathbf{k}$-point sampling of minimum $36\times 36$ for h-BN, $36\times 36$ for MoS$_2$ and $22\times 30$ for phosphorene. We also note that for large $\mathbf{k}$-point grids the band gaps using this method converge approximately as $1/N_\mathbf{k}$ as the missing contribution is almost proportional to the area of the $\mathbf{q} = \mathbf{0}$ region. Clearly, the latter approach varies significantly less with the $\mathbf{k}$-point grid and in fact the gap is converged to within \SI{0.2}{eV} already for a $\mathbf{k}$-point grid in the order of $6\times 6$ and to within $\sim\SI{0.1}{eV}$ with a $12\times 12$ grid (in the worst case). We have performed test calculations for other 2D semiconductors and obtained similar conclusions although the number of $\mathbf{k}$-points required to reach convergence within \SI{0.1}{eV} following the conventional approach ($\mathbf{q} = \mathbf{0}$ term neglected) is somewhat system dependent; materials with efficient screening, e.g. MoS$_2$ and NiS$_2$, require larger $\mathbf{k}$-point grids than materials with poor screening, e.g. h-BN and HfO$_2$ (see Fig.~\ref{fig:2d_eps}). 

\begin{figure}[ht]
  \includegraphics[width=\columnwidth]{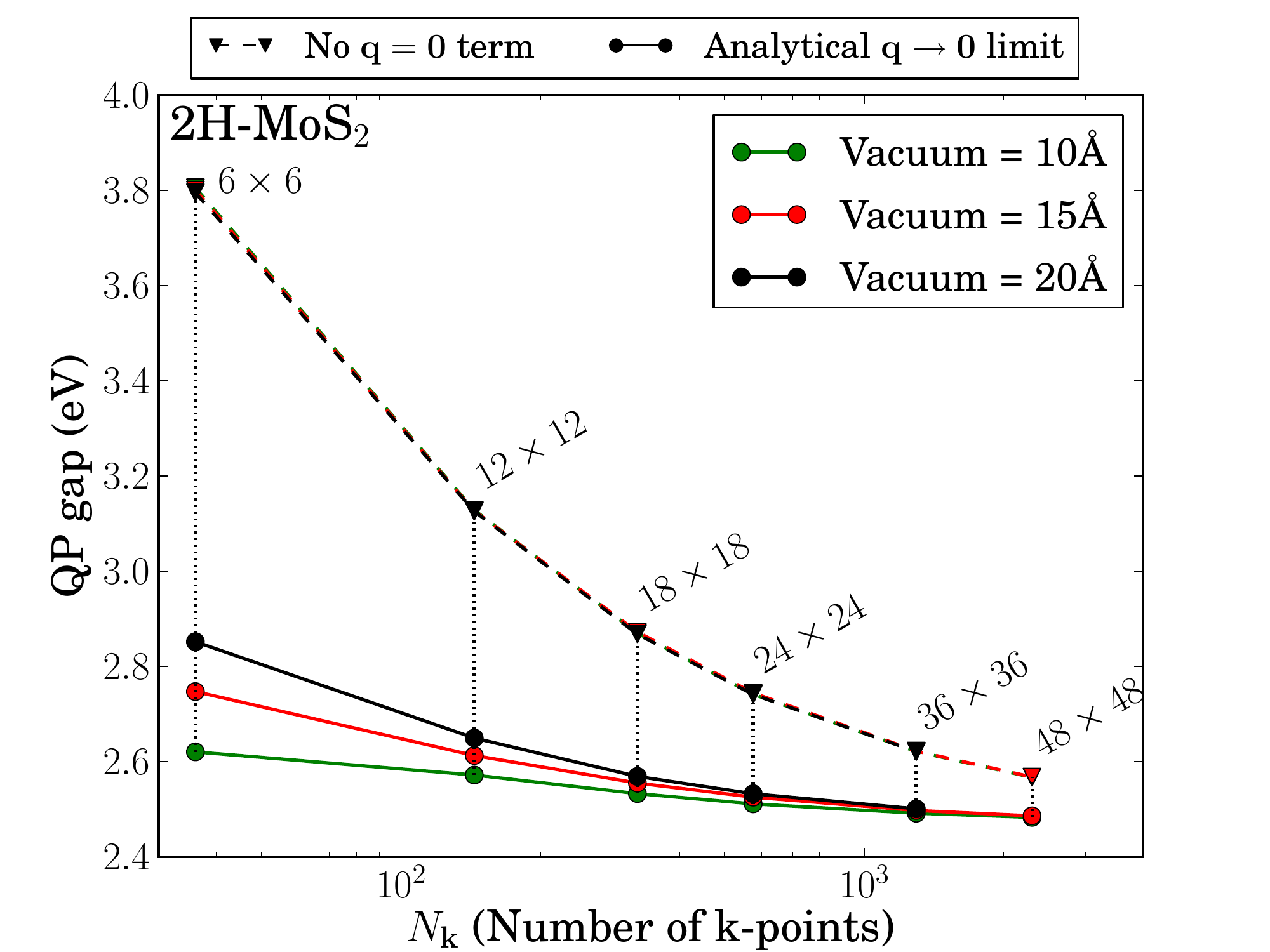}
  \caption{(Color online) The band gap of monolayer 2H-MoS$_2$ calculated with different amounts of vacuum between repeated layers. The solid and dashed lines are with and without the $\mathbf{q}=0$ correction, respectively. As the vacuum is increased, the weight of the correction is decreased and it is necessary to use denser in-plane $\mathbf{k}$-point sampling to achieve convergence.}
  \label{fig:kpt-vac}
\end{figure}

To obtain converged band gaps it is necessary to use a unit cell with enough vacuum between repeated layers to avoid an artificial interaction. This is true even when a truncated Coulomb interaction is used as the finite vacuum affects wave functions and energies, in particular for higher lying unbound states. As the amount of vacuum is increased, the Brillouin zone shrinks and the analytical correction around $\mathbf{q} = 0$, applied only for $\mathbf{G}=0$, has smaller weight. This means a slower convergence with respect to in-plane $\mathbf{k}$-points. This is shown in Fig.~\ref{fig:kpt-vac} for MoS$_2$, where it is clear that the correction is less effective for larger vacuum. The calculations converge toward the same value indicating that for MoS$_2$ \SI{10}{Å} of vacuum is sufficient. The most efficient procedure to obtain converged band gaps is therefore to first converge the amount of vacuum at a low $\mathbf{k}$-point sampling without applying the correction and then afterwards converge the $\mathbf{k}$-point sampling with the correction at the given vacuum.

In Table~\ref{tab:gaps} we report the converged values for the quasiparticle band gaps. For h-BN the band gap is indirect between the $K$- and $\Gamma$-point, for MoS$_2$ and phosphorene it is direct at the $K$- and $\Gamma$-point respectively. For these calculations we used $18\times 18$ $\mathbf{k}$-points for h-BN, $18\times 18$ $\mathbf{k}$-points for MoS$_2$ and $10\times 14$ for phosphorene with the analytical integration of $W(\mathbf{q})$ around $\mathbf{q} = \mathbf{0}$.  According to Fig.~\ref{fig:kpt-convergence} this is sufficient to ensure convergence to within \SI{0.05}{eV}. We note that spin-orbit interactions are not included in the reported values. Inclusion of spin-orbit interactions split the valence band of MoS$_2$ at the $K$ point by \SI{0.15}{eV} thereby lowering the QP gap by around \SI{0.07}{eV}\cite{zhu_giant_2011, rasmussen_computational_2015}. Spin-orbit interactions have no effect for h-BN and phosphorene. 

For MoS$_2$ the converged G$_0$W$_0$@PBE band gap of \SI{2.54}{eV} agrees well with our previously reported value of \SI{2.48}{eV} (with spin-orbit coupling) obtained using a Wigner-Seitz truncated Coulomb interaction and $30\times 30$ $\mathbf{k}$-points\cite{rasmussen_computational_2015}. Other reported gaps range from \SIrange{2.40}{2.82}{eV}\cite{ramasubramaniam_large_2012, chewchancham_2012, komsa_2012, molina_2013, Shi_2013, conley_bandgap_2013}.
However, these calculations were performed i) without the use of a truncated Coulomb interaction and including 15-25 Å vacuum, ii) employing relatively small $\mathbf{k}$-point grids of 6x6 to 18x18, and iii) using different in-plane lattice constants varying between 3.15 and 3.19 Å. These different settings can affect the band gap by as much as 0.5 eV \cite{huser_quasiparticle_2013}, and therefore we refrain from providing detailed comparison of our result to these earlier calculations. An overview of previous GW results for MoS$_2$ can be found in Ref.~\onlinecite{huser_quasiparticle_2013}.  

In Ref.~\onlinecite{qiu_erratum:_2015} a G$_0$W$_0$@LDA band gap for MoS$_2$ of \SI{2.70}{eV} is reported using a truncated Coulomb interaction and a calculation of the screened potential at $\mathbf{q} = \mathbf{0}$ based on the method in Ref.~\onlinecite{ismail-beigi_truncation_2006}. In that study, the lattice constant of MoS$_2$ was \SI{3.15}{Å}. With this lattice constant we obtain a gap of \SI{2.64}{eV}, which is in fair agreement with Ref.~\onlinecite{qiu_erratum:_2015}. Our result is very close to the experimental value of \SI{2.5}{eV} inferred from photo current spectroscopy\cite{klots_probing_2014}. Performing partially self-consistent GW$_0$ the band gap increases to \SI{2.65}{eV} (\SI{2.58}{eV} including spin-orbit coupling). 

For h-BN, we obtained a G$_0$W$_0$ band gap of \SI{7.06}{eV} which increases to \SI{7.49}{eV} with GW$_0$. In Ref.~\onlinecite{nieminen-hbn} the G$_0$W$_0$ band gap was calculated to be \SI{7.40}{eV}. Instead of a truncated Coulomb interaction the band gap was extrapolated to infinite vacuum. The treatment of the $\mathbf{q}=0$ term is not mentioned nor is the size of the $\mathbf{k}$-point grid. Despite the difference at the G$_0$W$_0$ level, they report a similar increase of the band gap of \SI{0.4}{eV} when doing a GW$_0$ calculation.

For phosphorene we calculate a G$_0$W$_0$ band gap of \SI{2.03}{eV} which agrees well with the previously reported value of \SI{2.0}{eV}\cite{tran_layer-controlled_2014} using the method of Ref.~\onlinecite{ismail-beigi_truncation_2006}. The band gap increases to \SI{2.29}{eV} with GW$_0$.

\begin{table}
\centering
\begin{ruledtabular}
  \begin{tabular}{lcccc}
     & Transition & DFT-PBE & G$_0$W$_0$@PBE & GW$_0$@PBE \\
     \cline{2-5}
     \noalign{\smallskip}
    h-BN & $K\rightarrow \Gamma$ & 4.64 & 7.06  & 7.49 \\
     &  $K\rightarrow K$ & 4.72 & 7.80 & 8.25 \\
    2H-MoS$_2$ & $K\rightarrow K$ & 1.65 & 2.54 & 2.65 \\
    Phosphorene & $\Gamma\rightarrow \Gamma$ & 0.90 & 2.03 & 2.29 \\
  \end{tabular}
\end{ruledtabular}
\caption{Band gaps in eV calculated with DFT-PBE, G$_0$W$_0$@PBE and GW$_0$@PBE using the PBE-relaxed structures. The GW calculations were performed using analytic integration of $\overline{W}(\mathbf{q})$ around $\mathbf{q} = \mathbf{0}$ without including spin-orbit interactions. \SI{10}{Å} of vacuum was used for MoS$_2$ and \SI{15}{Å} for h-BN and phosphorene. The following $\mathbf{k}$-point grids were used; h-BN: $18\times 18$, 2H-MoS2: $18\times 18$ and phosphorene: $10\times 14$.}
\label{tab:gaps}
\end{table}

\section{Conclusion}
In conclusion, we have discussed the connection between the form of the $\mathbf{q}$-dependent dielectric function of a 2D semiconductor and the slow $\mathbf{k}$-point convergence of the GW band structure. We have derived an analytical expression for the $\mathbf{q} \to \mathbf{0}$ limit of the screened potential of a semiconductor when a 2D truncation of the Coulomb potential is used. The method accounts for dielectric anisotropy and does not rely on any additional parameters or fitting. Using this expression we have shown that convergence of the GW self-energy with respect to the size of the $\mathbf{k}$-point grid is drastically improved. For the specific case of monolayer MoS$_2$, we found that the use of the analytical form alone reduces the $\mathbf{k}$-point grid required to achieve convergence of the GW self-energy contribution to the band gap to within $\sim\SI{0.1}{eV}$ from around $36\times 36$ to $12\times 12$ -- a reduction in the number of $\mathbf{k}$-points by an order of magnitude. This method may therefore enable future large-scale GW calculations for 2D materials without compromising accuracy. 

\begin{acknowledgments}
We acknowledge support from the Danish Council for Independent Research’s Sapere Aude Program, Grant No. 11-1051390. The Center for Nanostructured Graphene is sponsored by the Danish National Research Foundation, Project DNRF58.
\end{acknowledgments}

\appendix
\section{Calculation of the $\mathbf{q} \to \mathbf{0}$ limit of the screened potential}\label{sec:appendix}
In the following we derive the analytical form of the screened potential, Eq. \eqref{eq:screened-potential}, for 2D materials in the limit $\mathbf{q}_\parallel \to \mathbf{0}$. We largely follow the approach of Ref.~\onlinecite{freysoldt_dielectric_2007} where the same limit for bulk systems was considered. As explained in the main text we use a truncated Coulomb interaction of the form
\begin{equation}
  v(\mathbf{r}_\parallel, z) = \frac{\theta(R - |z|)}{\sqrt{|\mathbf{r}_\parallel|^2 + z^2}}.
\end{equation}
Using this potential we effectively turn off interaction between electrons on different 2D layers of the supercell calculation. We choose $R$ to be half the height of the unitcell, $R = L/2$, so that an electron in the center of the layer will not interact with electrons located in the neighboring unitcell. This means that the 2D truncated coulomb interaction of Eq. (16) in \cite{ismail-beigi_truncation_2006} reduces to
\begin{equation}
  v^\text{2D}_\mathbf{G}(\mathbf{q}_\parallel) = \frac{4\pi}{|\mathbf{q}_\parallel + \mathbf{G}|^2} \left[ 1 - e^{-|\mathbf{q}_\parallel + \mathbf{G}_\parallel| L / 2} \cos(|G_z| L / 2) \right], \label{eq:coulomb-2d}
\end{equation}
where only in-plane $\mathbf{q}$ is considered. We note that in the limit $L \to \infty$ it takes the usual 3D form, $v_\mathbf{G}(\mathbf{q}) = \frac{4\pi}{|\mathbf{q} + \mathbf{G}|^2}$. In the long wavelength limit it has the asymptotic behavior
\begin{align}
  v^\text{2D}_\mathbf{0}(\mathbf{q}_z = 0, \mathbf{q}_\parallel \to \mathbf{0}) = \frac{2\pi L }{|\mathbf{q}_\parallel|}, \label{eq:coulomb-2d-limit}
\end{align}
diverging slower than the full Coulomb potential with profound consequences for the properties of 2D materials.

In the long wavelength limit $\mathbf{q} \to \mathbf{0}$ the non-interacting density response function or irreducible polarizability has the following behavior\cite{yan_linear_2011} 
\begin{align}
  \chi^0_{\mathbf{00}'}(\mathbf{q} \to \mathbf{0}) ={}& \mathbf{q} \cdot \mathsf{P} \mathbf{q} = |\mathbf{q}|^2 \hat{\mathbf{q}} \cdot \mathsf{P} \hat{\mathbf{q}} \label{eq:chi-head}\\
  \chi^0_{\mathbf{G0}}(\mathbf{q}\to \mathbf{0}) ={}& \mathbf{q} \cdot \mathbf{p}_\mathbf{G} = |\mathbf{q}| \hat{\mathbf{q}} \cdot\mathbf{p}_\mathbf{G}, \label{eq:chi-wings-rows},\\
  \chi^0_{\mathbf{0G}}(\mathbf{q}\to \mathbf{0}) ={}& \mathbf{q} \cdot \mathbf{s}_\mathbf{G} = |\mathbf{q}| \hat{\mathbf{q}} \cdot\mathbf{s}_\mathbf{G}, \label{eq:chi-wings-cols}
\end{align}
where $\mathsf{P}$ is a second rank tensor, $\mathbf{p}_\mathbf{G}$ and $\mathbf{s}_\mathbf{G}$ are proper vectors and $\hat{\mathbf{q}} = \mathbf{q} / |\mathbf{q}|$. The density response function, and therefore also $\mathsf{P}$ and $\mathbf{p}_\mathbf{G}$, has a frequency dependence which here and through the rest of this section has been left out to simplify the notation.
Within the random phase approximation the dielectric function is given by (schematically)
\begin{equation}
  \varepsilon = 1 - v \chi^0.
\end{equation}
Due to technical reasons\cite{baldereschi_mean-value_1978,shishkin_implementation_2006} it is easier to work with a similar symmetrized version given in Fourier space by
\begin{equation}
  \tilde{\varepsilon}_{\mathbf{GG}'}(\mathbf{q}) = \delta_{\mathbf{GG}'} - \sqrt{v_\mathbf{G}(\mathbf{q})} \chi^0_{\mathbf{GG}'}(\mathbf{q}) \sqrt{v_{\mathbf{G}'}(\mathbf{q})}.
\end{equation}
Inserting the Coulomb potential, Eq. \eqref{eq:coulomb-2d}, and the expressions for the non-interacting response function Eqs. \eqref{eq:chi-head}-\eqref{eq:chi-wings-cols}, the head and wings of the symmetrized dielectric function are
\begin{align}
  \tilde{\varepsilon}_\mathbf{00}(\mathbf{q}_\parallel \to \mathbf{0}) ={}& 1 - v^\text{2D}_\mathbf{0}(\mathbf{q}_\parallel) |\mathbf{q}_\parallel|^2 \hat{\mathbf{q}_\parallel} \cdot \mathsf{P}\hat{\mathbf{q}_\parallel} \nonumber\\
  ={}& 1 - 4\pi (1 - e^{-|\mathbf{q}_\parallel| L/2}) \hat{\mathbf{q}_\parallel} \cdot \mathsf{P} \hat{\mathbf{q}_\parallel} \\
  \tilde{\varepsilon}_{\mathbf{G0}}(\mathbf{q}_\parallel \to \mathbf{0}) ={}& -\sqrt{v^\text{2D}_\mathbf{G}(\mathbf{0})} \sqrt{v_\mathbf{0}^\text{2D}(\mathbf{q}_\parallel)} \hat{\mathbf{q}_\parallel} \cdot \mathbf{p}_\mathbf{G} \nonumber\\
  ={}& -\sqrt{v^\text{2D}_\mathbf{G}(\mathbf{0})} \sqrt{4\pi (1 - e^{-|\mathbf{q}_\parallel| L/2})} \hat{\mathbf{q}_\parallel} \cdot \mathbf{p}_\mathbf{G}\\
  \tilde{\varepsilon}_{\mathbf{0G}}(\mathbf{q}_\parallel \to \mathbf{0}) ={}& -\sqrt{v^\text{2D}_\mathbf{G}(\mathbf{0})} \sqrt{v_\mathbf{0}^\text{2D}(\mathbf{q}_\parallel)} \hat{\mathbf{q}_\parallel} \cdot \mathbf{s}_\mathbf{G} \nonumber\\
  ={}& -\sqrt{v^\text{2D}_\mathbf{G}(\mathbf{0})} \sqrt{4\pi (1 - e^{-|\mathbf{q}_\parallel| L/2})} \hat{\mathbf{q}_\parallel} \cdot \mathbf{s}_\mathbf{G}.
\end{align}

To determine the inverse dielectric function we write the dielectric function as a block matrix in the $\mathbf{G}, \mathbf{G}'$ components with head, wings and body of the form
\begin{equation}
  \tilde{\boldsymbol{\varepsilon}} = \begin{pmatrix} H & \mathbf{w}^\intercal \\ \mathbf{v} & \mathbf{B} \end{pmatrix}
\end{equation}
The inverse is then given by
\begin{widetext}
\begin{equation}
  \tilde{\boldsymbol{\varepsilon}}^{-1} = \begin{pmatrix} (H - \mathbf{w}^\intercal \mathbf{B}^{-1} \mathbf{v})^{-1} & -(H - \mathbf{w}^\intercal \mathbf{B}^{-1} \mathbf{v})^{-1} \mathbf{w}^\intercal \mathbf{B}^{-1} \\ -\mathbf{B}^{-1} \mathbf{v} (H - \mathbf{w}^\intercal \mathbf{B}^{-1} \mathbf{v})^{-1} & \mathbf{B}^{-1} + \mathbf{B}^{-1} \mathbf{v} (H - \mathbf{w}^\intercal \mathbf{B}^{-1} \mathbf{v})^{-1} \mathbf{w}^\intercal \mathbf{B}^{-1} \end{pmatrix}
\end{equation}
From this we see that
\begin{align}
  \tilde{\varepsilon}^{-1}_\mathbf{00} ={}& \left[ \tilde{\varepsilon}_\mathbf{00} - \sum_{\mathbf{G}, \mathbf{G}' \neq \mathbf{0}} \tilde{\varepsilon}_\mathbf{0G} B^{-1}_{\mathbf{GG}'} \tilde{\varepsilon}_{\mathbf{G}'\mathbf{0}} \right]^{-1} \\
  \tilde{\varepsilon}^{-1}_\mathbf{G0} ={}& -\tilde{\varepsilon}^{-1}_\mathbf{00} \sum_{\mathbf{G}' \neq \mathbf{0}} B^{-1}_{\mathbf{GG}'} \tilde{\varepsilon}_{\mathbf{G}'\mathbf{0}} \\
\tilde{\varepsilon}^{-1}_\mathbf{0G} ={}& -\tilde{\varepsilon}^{-1}_\mathbf{00} \sum_{\mathbf{G}' \neq \mathbf{0}} \tilde{\varepsilon}_{\mathbf{0}\mathbf{G}'} B^{-1}_{\mathbf{G}'\mathbf{G}} \\
  \tilde{\varepsilon}^{-1}_{\mathbf{GG}'} ={}& B^{-1}_{\mathbf{GG}'} + \tilde{\varepsilon}^{-1}_{\mathbf{00}} \left( \sum_{\mathbf{G}'' \neq \mathbf{0}} B^{-1}_{\mathbf{GG}''} \tilde{\varepsilon}_{\mathbf{G}''\mathbf{0}} \right) \left( \sum_{\mathbf{G}'' \neq \mathbf{0}} \tilde{\varepsilon}_{\mathbf{0G}''} B^{-1}_{\mathbf{G}''\mathbf{G}'} \right)
\end{align}
\end{widetext}
Introducing the vectors $\mathbf{a}_\mathbf{G}$, $\mathbf{b}_\mathbf{G}$ and the tensor $\mathsf{A}$ given by
\begin{align}
  \mathbf{a}_\mathbf{G} ={}& -\sum_{\mathbf{G}' \neq \mathbf{0}} B^{-1}_{\mathbf{GG}'} \sqrt{v^\text{2D}_{\mathbf{G}'}(\mathbf{0})} \mathbf{p}_{\mathbf{G}'} \\
\mathbf{b}_\mathbf{G} ={}& -\sum_{\mathbf{G}' \neq \mathbf{0}} \sqrt{v^\text{2D}_{\mathbf{G}'}(\mathbf{0})} \mathbf{s}_{\mathbf{G}'} B^{-1}_{\mathbf{G}'\mathbf{G}} \\
  \mathsf{A} ={}& -\mathsf{P} - \sum_{\mathbf{G} \neq \mathbf{0}} \sqrt{v^\text{2D}_\mathbf{G}(\mathbf{q}_\parallel)} \mathbf{s}_\mathbf{G} \otimes \mathbf{a}_\mathbf{G},
\end{align}
where $\otimes$ denotes the tensor product, the long wavelength limit of the inverse dielectric function is seen to be given by
\begin{align}
  \tilde{\varepsilon}^{-1}_\mathbf{00}(\mathbf{q}_\parallel \to \mathbf{0}) ={}& \frac{1}{1 + 4\pi (1 - e^{-|\mathbf{q}_\parallel| L/2}) \hat{\mathbf{q}_\parallel} \cdot \mathsf{A}\hat{\mathbf{q}_\parallel}} \\
  \tilde{\varepsilon}^{-1}_{\mathbf{G0}}(\mathbf{q}_\parallel \to \mathbf{0}) ={}& -\frac{\sqrt{4\pi (1 - e^{-|\mathbf{q}_\parallel| L/2})} \hat{\mathbf{q}_\parallel} \cdot \mathbf{a}_\mathbf{G}}{1 + 4\pi (1 - e^{-|\mathbf{q}_\parallel| L/2}) \hat{\mathbf{q}_\parallel} \cdot \mathsf{A} \hat{\mathbf{q}_\parallel}} \\
  \tilde{\varepsilon}^{-1}_{\mathbf{0G}}(\mathbf{q}_\parallel \to \mathbf{0}) ={}& -\frac{\sqrt{4\pi (1 - e^{-|\mathbf{q}_\parallel| L/2})} \hat{\mathbf{q}_\parallel} \cdot \mathbf{b}_\mathbf{G}}{1 + 4\pi (1 - e^{-|\mathbf{q}_\parallel| L/2}) \hat{\mathbf{q}_\parallel} \cdot \mathsf{A} \hat{\mathbf{q}_\parallel}} \\
  \tilde{\varepsilon}^{-1}_{\mathbf{GG}'}(\mathbf{q}_\parallel \to \mathbf{0}) ={}& B^{-1}_{\mathbf{GG}'} \nonumber\\
 & + \frac{4\pi (1 - e^{-|\mathbf{q}_\parallel| L/2}) (\hat{\mathbf{q}_\parallel} \cdot \mathbf{a}_\mathbf{G}) (\hat{\mathbf{q}_\parallel} \cdot \mathbf{b}_{\mathbf{G}'})}{1 + 4\pi (1 - e^{-|\mathbf{q}_\parallel| L/2}) \hat{\mathbf{q}_\parallel} \cdot \mathsf{A} \hat{\mathbf{q}_\parallel}}
\end{align}

Inserting these expression in the equation for the screened potential, Eq. \eqref{eq:screened-potential}, we see that the head and wings are given by
\begin{widetext}
\begin{align}
  \overline{W}_\mathbf{00}(\mathbf{q}_\parallel \to \mathbf{0}) ={}& v^\text{2D}_\mathbf{0}(\mathbf{q}_\parallel) \left[ \tilde{\varepsilon}^{-1}_\mathbf{00}(\mathbf{q}_\parallel) - 1\right] \nonumber\\
  ={}& -\left( \frac{4\pi (1 - e^{-|\mathbf{q}_\parallel| L/2})}{|\mathbf{q}_\parallel|} \right)^2  \frac{\hat{\mathbf{q}_\parallel} \cdot \mathsf{A}\hat{\mathbf{q}_\parallel}}{1 + 4\pi (1 - e^{-|\mathbf{q}_\parallel| L/2}) \hat{\mathbf{q}_\parallel} \cdot \mathsf{A} \hat{\mathbf{q}_\parallel}} \label{eq:Wheadfull}\\
  \overline{W}_\mathbf{G0}(\mathbf{q}_\parallel \to \mathbf{0}) ={}& \sqrt{v^\text{2D}_\mathbf{G}(\mathbf{0})} \tilde{\varepsilon}^{-1}_\mathbf{G0}(\mathbf{q}_\parallel) \sqrt{v^\text{2D}_\mathbf{0}(\mathbf{q}_\parallel)} \nonumber\\
  ={}& -\frac{4\pi (1 - e^{-|\mathbf{q}_\parallel| L/2})}{|\mathbf{q}_\parallel|} \frac{\sqrt{v^\text{2D}_\mathbf{G}(\mathbf{0})} \hat{\mathbf{q}_\parallel} \cdot \mathbf{a}_\mathbf{G}}{1 + 4\pi (1 - e^{-|\mathbf{q}_\parallel| L/2}) \hat{\mathbf{q}_\parallel} \cdot \mathsf{A} \hat{\mathbf{q}_\parallel}} \\
  \overline{W}_\mathbf{0G}(\mathbf{q}_\parallel \to \mathbf{0}) ={}& \sqrt{v^\text{2D}_\mathbf{G}(\mathbf{0})} \tilde{\varepsilon}^{-1}_\mathbf{G0}(\mathbf{q}_\parallel) \sqrt{v^\text{2D}_\mathbf{0}(\mathbf{q}_\parallel)} \nonumber\\
  ={}& -\frac{4\pi (1 - e^{-|\mathbf{q}_\parallel| L/2})}{|\mathbf{q}_\parallel|} \frac{\sqrt{v^\text{2D}_\mathbf{G}(\mathbf{0})} \hat{\mathbf{q}_\parallel} \cdot \mathbf{b}_\mathbf{G}}{1 + 4\pi (1 - e^{-|\mathbf{q}_\parallel| L/2}) \hat{\mathbf{q}_\parallel} \cdot \mathsf{A} \hat{\mathbf{q}_\parallel}}.
\end{align}
and the body also gets a correction and becomes
\begin{align}
  \overline{W}_{\mathbf{GG}'}(\mathbf{q}_\parallel\to\mathbf{0}) ={}& \sqrt{v^\text{2D}_\mathbf{G}(\mathbf{0}) v^\text{2D}_{\mathbf{G}'}(\mathbf{0})} \left[\varepsilon^{-1}_{\mathbf{GG}'}(\mathbf{q}_\parallel) - \delta_{\mathbf{GG}'}\right] \nonumber\\
  \begin{split} ={}& \sqrt{v^\text{2D}_\mathbf{G}(\mathbf{0}) v^\text{2D}_{\mathbf{G}'}(\mathbf{0})} \left[ B^{-1}_{\mathbf{GG}'} - \delta_{\mathbf{GG}'} + \frac{4\pi (1 - e^{-|\mathbf{q}_\parallel| L/2}) (\hat{\mathbf{q}_\parallel} \cdot \mathbf{a}_\mathbf{G}) (\hat{\mathbf{q}_\parallel} \cdot \mathbf{b}_{\mathbf{G}'})}{1 + 4\pi (1 - e^{-|\mathbf{q}_\parallel| L/2}) \hat{\mathbf{q}_\parallel} \cdot \mathsf{A} \hat{\mathbf{q}_\parallel}} \right]. \label{eq:Wbodyfull}
  \end{split}
\end{align}
\end{widetext}
Taking the limit $\mathbf{q}_\parallel \to \mathbf{0}$ we see that the head of the screened potential is $\overline{W}_\mathbf{00}(\mathbf{q}_\parallel) \to \mathbf{0}) = (2\pi L)^2 \hat{\mathbf{q}_\parallel} \cdot \mathsf{A}\hat{\mathbf{q}_\parallel}$ which is a finite value.

Defining the dimensionless quantity $\mathbf{x} = \mathbf{q}_\parallel L/2$ and the rotational average $A = \frac{1}{2\pi}\int_0^{2\pi}\hat{\mathbf{x}}(\phi) \cdot \mathsf{A} \hat{\mathbf{x}}(\phi) d\phi$ it is possible to rewrite the head of the screened potential as
\begin{align} \label{eq:w_dimless}
\tilde{w}(\mathbf{x}) &= \frac{\overline{W}(2\mathbf{x}/L)}{(2\pi L)^2 A} \nonumber \\
&= - \left( \frac{1-e^{-|\mathbf{x}|}}{|\mathbf{x}|} \right)^2 \frac{1}{1+4\pi A (1-e^{-|\mathbf{x}|})}.
\end{align}
It is evident that the polar integral of Eq.~(\ref{eq:w_dimless}), over a small circle with radius $r_{\Omega_0}$, cannot be evaluated analytically:
\begin{align}
&\int_{\Omega_0} \tilde{w}(\mathbf{x})\,\mathrm{d}\mathbf{x} = \nonumber \\
& -2\pi \int_0^{r_{\Omega_0}} \left( \frac{1-e^{-\mathbf{x}}}{\mathbf{x}} \right)^2 \frac{1}{1+4\pi A (1-e^{-\mathbf{x}})} \mathbf{x}\,\mathrm{d}\mathbf{x}.
\end{align}
It is however noticed that the function $\tilde{y}(\mathbf{x}) = \frac{1}{1+(1+4\pi A)\mathbf{x}}$ agrees very well with the integrand for small $\mathbf{x}$. It has the same first order Taylor expansion and it is integrable. This yields
\begin{align} \label{eq:w_taylor}
&\int_{\Omega_0} \tilde{w}(\mathbf{x})\,\mathrm{d}\mathbf{x} \approx 
 -2\pi \int_0^{r_{\Omega_0}}  \frac{\mathbf{x}}{1+(1+4\pi A)\mathbf{x}} \,\mathrm{d}\mathbf{x} = \nonumber \\
& \frac{-2\pi (4\pi A r_{\Omega_0} + r_{\Omega_0} - \ln(4\pi A r_{\Omega_0} + r_{\Omega_0} + 1)}{(4\pi A + 1)^2}.
\end{align}

Since the expression in Eq.~(\ref{eq:w_taylor}) only holds for small $\mathbf{x}$, it is generally not valid in the entire $\Omega_0$ region. Also, the expression is not valid for non-isotropic materials, where it is not justified to take the rotational average of the $\mathsf{A}$ tensor.

\bibliography{references}

\end{document}